\documentclass[english]{article}
\usepackage[T1]{fontenc}
\usepackage[latin9]{inputenc}
\usepackage{geometry}
\usepackage{color}
\usepackage{amsmath}
\usepackage{amsthm}
\usepackage{wasysym}

\makeatletter
\theoremstyle{plain}
\newtheorem{thm}{\protect\theoremname}
\theoremstyle{remark}
\newtheorem{rem}[thm]{\protect\remarkname}

\usepackage{babel}

\makeatother

\usepackage{babel}
\providecommand{\remarkname}{Remark}
\providecommand{\theoremname}{Theorem}

\begin{document}
\title{\textbf{Response to a critique of ``Cotton Gravity''}}
\author{R.A. Sussman$^{1}$, C.A. Mantica $^{2}$, L.G. Molinari $^{2}$ and
S. Najera $^{1}$}
\date{$^{1}$Institute of Nuclear Sciences, National Autonomous University
of Mexico, UNAM\\
 $^{2}$Physics Department Aldo Pontremoli, Università degli Studi
di Milano \\
 and I.N.F.N. sezione di Milano\\[2ex]
}
\maketitle
\begin{abstract}
We address in this article the criticism in a recently submitted article
by Clement and Noiucer \texttt{ArXiv:2312.17662 {[}gr-qc{]}} on ``Cotton
Gravity'' (CG), a gravity theory alternative to General Relativity.
These authors claim that CG is ``not predictive'' for producing
``too many" spherically symmetric vacuum solutions, while taking
the Bianchi I vacuum as test case they argue that geometric constraint
on the Cotton tensor lead to an undetermined problem, concluding in
the end that CG ``is not a physical theory''. We provide arguments
showing that this critique is incorrect and misrepresents the theory. 
\end{abstract}

\section{Introduction: ``Cotton'' gravity vs ``Codazzi'' gravity}

The field equations of Cotton Gravity (denoted by ``CG'') derived
originally by J. Harada are \cite{Harada1,Harada2} 
\begin{eqnarray}
C_{abc} & = & 16\pi{\cal T}_{abc},\label{CottonFeqs}\\
\hbox{where}\quad{\cal T}_{abc} & = & \nabla_{a}T_{bc}-\nabla_{b}T_{ac}-\frac{1}{3}\left(g_{bc}\nabla_{a}T-g_{ac}\nabla_{b}T\right),\label{Tcot}
\end{eqnarray}
and $C_{abc}$ is the Cotton tensor given by \cite{CottonWeyl} 
\begin{equation}
C_{abc}\equiv\nabla_{a}R_{bc}-\nabla_{b}R_{ac}-\frac{1}{6}\left(g_{bc}\nabla_{a}R-g_{ac}\nabla_{b}R\right),\label{Cotton}
\end{equation}
Clement and Noiucer present in \cite{Clement} two main arguments:
(i) the alleged existence of an infinite number of spherically symmetric
static vacuum solutions (incompatibility with Birkhoff's theorem)
and (ii) restrictions posed by geometric constraints on the Cotton
tensor, in particular $g^{bc}C_{abc}=0$, which in their view limits
the space of solutions of Bianchi I vacuum spacetimes. Notice that
these seem to be contradictory claims, as (1) might lead to overdetermination
(too many solutions) and (2) to underdetermination. Further, Clement
and Noiucer argue that the shortcomings they claim to have found in
CG become more severe as the theory is applied to spacetimes with
weaker symmetries.

Clement and Noiucer only considered the field equations in the form
\eqref{CottonFeqs}-\eqref{Cotton}, but there is an alternative formulation
of these field equations, given in terms of a Codazzi tensor, derived
by Mantica and Molinari \cite{Mantica1,Mantica2,Mantica3} and used
by Sussman and Najera \cite{Sussman1,Sussman2}: 
\begin{eqnarray}
 & {} & \nabla_{b}{\cal C}_{ac}-\nabla_{c}{\cal C}_{ab}=0,\qquad{\cal C}_{ac}\;\text{must be a Codazzi tensor,}\label{Codazzi}\\
 & {} & \hbox{where}\quad{\cal C}_{ab}=G_{ab}-8\pi T_{ab}-\frac{1}{3}(G-8\pi T)g_{ab},\label{preCodazzi}
\end{eqnarray}
with $G_{ab},\,\,T_{ab}$ the Einstein and energy momentum tensors
and $G,\,\,T$ their traces. Clement and Noiucer basically disregard
this formulation, arguing that it carries the shortcomings they claim
affect the field equations \eqref{CottonFeqs}-\eqref{Cotton}, a
claim we disprove in this response.\\

As we show in this response, the alternative Codazzi formulation to
CG provides a much deeper theoretical insight into CG than the Cotton
formulation of \eqref{CottonFeqs}-\eqref{Cotton}. Their dismissal
of the Codazzi approach \eqref{Codazzi}- \eqref{preCodazzi} (at
least) partly explains the misconceptions of the criticism by Clement
and Noiucer. The main advantages of the formulation furnished by \eqref{Codazzi}-\eqref{preCodazzi}
are: 
\begin{itemize}
\item It does not rely on the 3rd order tensor ${\cal T}_{abc}$ in \eqref{Tcot}
associated with a generalized angular momentum tensor \cite{Gogber}.
However, in the field equations \eqref{CottonFeqs} it appears to
play the role of a general matter-energy source for CG, which is deeply
misleading because it is clearly not applicable universally as a matter-energy
source, as for example the role of $T_{ab}$ in GR. While ${\cal T}_{abc}$
is constructed from the physical source $T_{ab}$, it is either impossible
or very hard to find $T_{ab}$ by deconstructing ${\cal T}_{abc}$
when using the Cotton field equations \eqref{CottonFeqs}. Also,
the dependence of ${\cal T}_{abc}$ on derivatives of the metric (from
the covariant derivatives of $T_{ab}$) introduces multiple ambiguities
that complicate its interpretation. For example, for the same $T_{ab}$
(say, dust or scalar field) for two distinct spacetimes will necessarily
yield very different (likely unrelated) forms of ${\cal T}_{abc}$,
which might loose all physical information contained in $T_{ab}$.

Thus, given the limitations and ambiguities of ${\cal T}_{abc}$,
setting ${\cal T}_{abc}=0$ as the condition for a vacuum spacetime
leads to confusion (as seen in the article by Clement and Noiucer).
As a contrast, the Codazzi formulation \eqref{Codazzi}-\eqref{preCodazzi}
relies directly on $T_{ab}$ without using ${\cal T}_{abc}$. This
allows for a much higher degree of control and consistency over $T_{ab}$,
which is a universally applicable tensor to describe field sources.
In particular, it is useful to select beforehand (at \eqref{preCodazzi})
the same $T_{ab}$ used in GR solutions when trying to find their
generalization in CG.
\item It avoids the failure of the field equations \eqref{CottonFeqs}-\eqref{Cotton}
to handle non-vacuum conformally flat spacetimes (such as FLRW models),
since it does not require computing beforehand the Cotton tensor \eqref{Cotton}
that vanishes automatically when the Weyl tensor vanishes \cite{CottonWeyl}.
Clement and Noiucer claim (it appears in their abstract) that FLRW
models are trivial solutions of CG, but non-trivial solutions for
these models (and any non-vacuum conformally flat spacetime) can be
perfectly handled with \eqref{Codazzi}-\eqref{preCodazzi}. 
\item It reduces the third order problem in the field equations \eqref{CottonFeqs}-\eqref{Cotton}
to a ``2+1'' order problem in two stages: first constructing a nonzero
second order tensor ${\cal C}_{ab}$ using Einstein's equations as
reference in \eqref{preCodazzi}, completing an extra order of differentiation
in \eqref{Codazzi}. In all the process, the source is $T_{ab}$. 
\item It facilitates finding solutions of the field equations that generalize
known GR solutions fulfilling the Correspondence Principle, which
is an important requirement in alternative gravity theories \cite{Ellis}. 
\end{itemize}
As we show further ahead, these features of the Codazzi formulation
eliminate most of the ambiguities mentioned by Clement and Noiucer,
which originate from their exclusive usage of the Cotton formulation
\eqref{CottonFeqs}-\eqref{Cotton}. In fact, the advantages of these
field equations fully justifies renaming the theory as ``Codazzi
Gravity'' instead of ``Cotton Gravity''. In what follows we examine
the main points of criticism raised by Clement and Noiucer.

\section{Vacuum solutions in static spherical symmetry}

\label{VacSols}

Clement and Noiucer claim that CG generates an infinite number of
vacuum solutions for static spherical symmetry. We prove in this section
that this claim is false. These authors considered the spherically
symmetric static metric 
\begin{equation}
ds^{2}=-A(r)dt^{2}+\frac{dr^{2}}{B(r)}+r^{2}(d\theta^{2}+\sin^{2}\theta d\phi^{2}),\label{staticSS}
\end{equation}
which we write in a different notation (it relates to theirs by $A=e^{2\lambda},\,\,B=e^{-2\mu}$).
They assumed that vacuum solutions of CG (as stated in \eqref{CottonFeqs}-\eqref{Cotton})
follow from the condition 
\begin{equation}
{\cal T}_{abc}=\nabla_{a}T_{bc}-\nabla_{b}T_{ac}-\frac{1}{3}\left(g_{bc}\nabla_{a}T-g_{ac}\nabla_{b}T\right)=0.\label{vacuumC}
\end{equation}
which clearly implies $C_{abc}=0$ from \eqref{Cotton}. They obtain
for \eqref{staticSS} a single third order differential equation (their
equation (13)) from $C_{abc}=0$, which in our notation reads: 
\begin{eqnarray}
\frac{A'''}{A}+\left(\frac{3B'}{2B}-\frac{2A'}{A}+\frac{2}{r}\right)\frac{A''}{A}+\left(\frac{A'}{A}-\frac{B'}{B}-\frac{1}{2r}\right)\frac{A'^{2}}{A^{2}}+\left(\frac{B''}{2B}+\frac{B'}{2rB}-\frac{2}{r^{2}}\right)\frac{A'}{A}+\frac{2}{r^{2}}\left(1-\frac{1}{B}\right)=0.\nonumber \\
\label{Cottzero}
\end{eqnarray}
This complicated differential equation involves a non-linear coupling
of the two metric coefficients and thus admits an infinite number
of solutions with (in general) 5 integration constants. As an example,
one solution of \eqref{Cottzero} emerges for each one in the infinite
possible choice of either one of $A$ or $B$ (or other restrictions).
But are all these solutions vacuum solutions of CG for \eqref{staticSS}?

Moreover, Clement and Noiucer made no assumption on $T_{ab}$ when
computing $C_{abc}=0$, so they perhaps implicitly assumed that ${\cal T}_{abc}=0$
describes vacuum solutions for an arbitrary $T_{ab}$.

Clearly (in fact, trivially) $T_{ab}=0\,\,\Rightarrow\,\,T_{abc}=0$
in \eqref{vacuumC}, but this does not imply that ${\cal T}_{abc}=0$
ensures that $T_{ab}=0$ holds, which is not a trivial matter.\\

\noindent \textbf{Proposition.} The statement ${\cal T}_{abc}=0\,\,\Rightarrow\,\,T_{ab}=0$
is in general false for static spherical symmetry. \\

\noindent \textbf{Proof.} Consider the general case of the metric
\eqref{staticSS} with $A\ne B$. This metric in a static comoving
frame ($u^{a}=A^{-1/2}\delta_{t}^{a}$) admits as the most general
energy momentum tensor a fluid with anisotropic pressure 
\begin{eqnarray}
T_{ab} & = & \rho u_{a}u_{b}+ph_{ab}+\Pi_{ab},\qquad\Pi_{a}^{a}=0,\quad\Pi_{r}^{r}=-2P,\,\,\Pi_{\theta}^{\theta}=\Pi_{\phi}^{\phi}=P,\,\,\Pi_{t}^{t}=0,\label{TabST}\\
\nabla_{b}T^{ab} & = & 0\quad\Rightarrow\quad(p-2P)'=-\frac{\rho+p-2P}{2}\frac{A'}{A}+\frac{3P}{r}.\label{conST}
\end{eqnarray}
where a prime denotes derivative with respect to $r$, and $\rho,\,p,\,P$
are functions of $r$. Inserting \eqref{TabST}-\eqref{conST} into
the condition ${\cal T}_{abc}=0$ with ${\cal T}_{abc}$ given by
\eqref{vacuumC} leads after simple but laborious algebraic manipulation
to the following constraints 
\begin{equation}
p'=-\frac{[\rho+p-2P]\,A'}{2A}-\frac{2}{3}\rho',\qquad P'=-\frac{1}{3}\rho'-\frac{3P}{r},\label{constrST1}
\end{equation}
which together with \eqref{conST} can be combined into 
\begin{equation}
P'-\frac{3}{2}\frac{P}{r}-\frac{1}{3}\rho'=0.\label{constrST2}
\end{equation}
Therefore, condition $C_{abc}=0$ in the field equations \eqref{CottonFeqs}
(which follows from ${\cal T}_{abc}=0$) does not lead univocally
to vacuum CG solutions, as it is clearly compatible with a large class
of non-vacuum energy momentum tensors \eqref{TabST} with $p,\,P,\,\rho$
linked by \eqref{constrST1}-\eqref{constrST2}. As a consequence,
Clement and Noiucer have not proven that CG has inherently an infinite
number of vacuum spherical solutions from the condition $C_{abc}=0$.

The failure of $C_{abc}=0$ to lead to a vacuum solution for \eqref{staticSS}
comes from the usage of the Cotton formulation of CG in \eqref{CottonFeqs},
since ${\cal T}_{abc}=0$ provides no information on $T_{ab}$ and
might be very different for the same $T_{ab}$ in different spacetimes.
It is practically impossible using \eqref{CottonFeqs}-\eqref{Cotton}
to consistently associate to the solutions of $C_{abc}=0$ their corresponding
$T_{ab}$ without actually finding the solutions explicitly.

However, this ambiguity can be resolved using the Correspondence Principle
to GR, together with the Codazzi formulation \eqref{Codazzi}-\eqref{preCodazzi},
which only contains $T_{ab}$ and can verify its consistency with
reference to a correspondence with Einstein's equations. We illustrate
this process with a vacuum and non-vacuum solution to $C_{abc}=0$
using \eqref{Codazzi}-\eqref{preCodazzi}. 
\begin{description}
\item [{A unique vacuum solution}] A spherically symmetric vacuum solution
in GR should describe a point mass, thus we expect a CG solution to
relate also to a point mass. In the Codazzi formulation this solution
follows by substituting the energy momentum tensor \eqref{TabST}
into $T_{ab}-\frac{1}{3}Tg_{ab}=0$ which implies $\rho-p=P=0$, a
true vacuum state. Then \eqref{Codazzi}-\eqref{preCodazzi} take
the form %
\begin{equation}
\nabla_{c}{\cal C}_{ab}-\nabla_{b}{\cal C}_{ac}=0\quad\hbox{for}\quad{\cal C}_{ab}=G_{ab}-\frac{1}{3}G\,g_{ab},\label{CGvacuumSS}
\end{equation}
Since ${\cal C}_{ab}$ is closely linked to $G_{ab}$, we need to
bear in mind (as reference) how the unique vacuum solution of GR for
static spherical symmetry is derived. A necessary (not sufficient)
condition for such a solution is $A=B$, but the general solution
follows from $G_{ab}-Gg_{ab}=0$ are the Schwarzschild metric functions
$A=B=1-2M_{s}/r$ with $M_{s}$ being the Schwarzschild mass. Evidently,
substitution of these metric functions of GR leads (by construction)
to ${\cal C}_{ab}=0$. However, if $A\ne B$, then the vacuum condition
$T_{ab}-\frac{1}{3}Tg_{ab}$ does not hold, so $T_{ab}\ne0$.

To obtain then a non-trivial CG vacuum solution with a consistent
correspondence with GR, we assume the most general possible deviation
from Schwarzschild that satisfies the necessary condition $A=B$ 
\begin{equation}
A=B=\Phi(r),\label{CGtrial1}
\end{equation}
where $\Phi(r)$ is an arbitrary function that will be determined
by \eqref{CGvacuumSS}. Inserting \eqref{CGtrial1} into \eqref{CGvacuumSS}
leads to a nonzero ${\cal C}_{ab}$ 
\begin{equation}
C_{\theta}^{\theta}=C_{\phi}^{\phi}=-2C_{t}^{t}=-2C_{r}^{r}=\frac{\Phi_{,rr}r^{2}-2\Phi_{,r}r-4\Phi+1}{6r^{2}},\label{Crn}
\end{equation}
which inserted in \eqref{Codazzi} leads to the following linear third
order differential equation 
\begin{equation}
r^{3}\Phi_{,rrr}+r^{2}\Phi_{,rr}-2r\Phi_{,r}+2\Phi-2=0,
\end{equation}
whose unique solution is 
\begin{equation}
\Phi(r)=1-\frac{r_{s}}{r}+\gamma\,r+\frac{8\pi}{3}\,\Lambda\,r^{2}.\label{Schw2}
\end{equation}
where we have identified the integration constants as the Schwarzschild
radius $r_{s}$ and the cosmological constant (emerging naturally
as an integration constant), while $\gamma r$ is the linear term
inherent of CG. This is the solution found by Harada in \cite{Harada1}

This is indeed a vacuum solution of CG for \eqref{staticSS} because
it fulfills the criterion that defines a vacuum state $T_{ab}-\frac{1}{3}Tg_{ab}=0$.
Also, it is a unique vacuum solution, since $T_{ab}-\frac{1}{3}Tg_{ab}\ne0$
for $A\ne B$. It exhibits a clear correspondence with the GR vacuum
solution (Schwarzschild follows from $\gamma=\lambda=0$) and with
the Schwarzschild-Kottler solution (setting $\gamma=0$).
\item [{Non-vacuum solutions.}] ~

The key question now is if there are other CG vacuum solutions for
\eqref{staticSS}, which must comply with solving \eqref{Cottzero}
with $A\ne B$, which is a technically difficult problem. Even assuming
only correspondence with GR by choosing $A=1-2M_{s}/r+\alpha(r)$
and $B=1-2M_{s}/r+\beta(r)$ with $\alpha(r),\,\,\beta(r)$ arbitrary
(including polynomials) leads to a third differential equation as
formidable as \eqref{Cottzero}.

Solutions with $A\ne B$ are not vacuum solutions, since $T_{ab}-\frac{1}{3}Tg_{ab}\ne0$.
In this case instead of \eqref{CGvacuumSS} we have to use the general
form of \eqref{Codazzi} and \eqref{preCodazzi}, which is equivalent
to \eqref{CottonFeqs} with ${\cal T}_{abc}=0$, but with $T_{ab}\ne0$
given by \eqref{TabST}-\eqref{constrST2}. In fact, as we showed
before, $\rho=p=P=0$ in \eqref{TabST}-\eqref{constrST2} holds only
if $T_{ab}-\frac{1}{3}Tg_{ab}=0$ holds.

It is illustrative to test solutions of \eqref{Cottzero} to verify
by correspondence with GR if they comply with the conditions for being
vacuum solutions. For example, if we set $B=CA$, with $C(r)$ an
arbitrary function, \eqref{Cottzero} becomes linear by eliminating
the coefficient of $A'^{2}/A^{2}$, leading to a simple analytic solution
with $A\ne B$ 
\begin{equation}
B=\frac{C_{0}A}{\sqrt{r}},\qquad A=C_{1}r^{5/4}+\frac{C_{2}}{\sqrt{r}}-C_{3}\frac{8r^{2}}{15}+\frac{16\sqrt{r}}{9}.\label{nonvac1}
\end{equation}
where $C_{0},\,C_{1},\,C_{2},\,C_{3}$ are integration constants.
To find out an energy momentum tensor that can be associated with
\eqref{nonvac1}, we insert these metric functions into $G_{ab}$,
leading to 
\begin{equation}
8\pi\rho=u^{a}u^{b}G_{ab}=\frac{1}{\sqrt{r}}\left[\frac{4}{3}c_{0}c_{4}+\frac{1}{r^{3/2}}\left(1-\frac{16c_{0}}{9}\right)-\frac{7c_{0}c_{1}}{4r^{1/4}}\right]
\end{equation}
with $8\pi p=(1/3)h_{ab}G^{ab}$ and $8\pi\Pi^{ab}=[h_{(c}^{a}h_{d)}^{b}-(1/3)h_{cd}]G^{cd}$
given by long expressions that we omit. Contracting these forms in
the same way with ${\cal C}_{ab}=G_{ab}-\frac{1}{3}Gg_{ab}$ leads
to similar results. Therefore, \eqref{nonvac1} is not a vacuum solution
and bears no relation to a point mass solution.\\
 Gogberashvili and Girgvliani \cite{Gogber} found exact solutions
by fixing the metric function $A$ in \eqref{staticSS} and proposed
physical interpretations, but none of these solutions is a vacuum
solution and none bears any relation with a simple point mass solution.
\item [{Birkhoff's theorem.}] The arguments presented above seem to suggest
that \eqref{Schw2} is the unique CG vacuum solution for static spherical
symmetry. This seems to suggest a high likelihood that CG with the
Codazzi formulation is compatible with Birkhoff's theorem, though
we still need a more rigorous proof (this is left for future work).
However, the lack of fulfillment of Birkhoff's theorem seems to be
a feature common to $f(R)$ theories \cite{fR1,fR2,Capozziello 07}.
The authors of \cite{fR1} found an infinite number of static spherically
symmetric vacuum solutions of a class of $f(R)$ theories depending
on an arbitrary function (see their eq 16).

Nevertheless, we have proved that the claim by Clement and Noiucer
is mistaken and given the complexity of \eqref{Cottzero} with $A\ne B$
and the simplicity of a point mass, it seems extremely unlikely that
\eqref{Cottzero} admits other vacuum solutions. though at least it
admits \eqref{Schw2} as a point mass solution with proper correspondence
to GR. 
\end{description}

\subsection{The Bianchi I vacuum}

The second case example used by Clement and Noiucer to assess CG is
the vacuum Bianchi I spacetime (see Section 13.3 in \cite{Stephani}).
They argue that the condition $g^{bc}C_{abc}=0$ leads to an underdetermination
restricting the set of CG solutions, since one of the 3 nonzero component
of $C_{abc}=0$ can be eliminated in terms of the other two (as opposed
to the ``too many'' solutions in spherically symmetric vacuum).
We show that no such restriction arises when looking at vacuum solutions
of the Bianchi I models under the Codazzi formulation \eqref{Codazzi}-\eqref{preCodazzi}.
However, using $C_{abc}=0$ with the Cotton formulation \eqref{CottonFeqs}-\eqref{Cotton}
leads to the same problem as in the spherical static case, namely:
condition $T_{abc}=0$ does not imply $T_{ab}=0$, thus it is not
certain that the arguments of Clement and Noiucer are relevant for
Bianchi I vacuum.

Consider the vacuum Bianchi I model described by the metric 
\begin{equation}
ds^{2}=-dt^{2}+A^{2}(t)dx^{2}+B^{2}(t)dy^{2}+C^{2}(t)dz^{2},\label{Bianchi1}
\end{equation}
whose GR solution follows from $R_{ab}=0$ (or $G_{ab}=0$) using
the Kasner parametrization $A=t^{m},\,\,B=t^{n},\,\,C=t^{p}$, where
$m,\,n,\,p$ are real constants \cite{Stephani}. Substitution in
$R_{ab}=0$ yields the following constraints 
\begin{equation}
R_{x}^{x}=R_{y}^{y}=R_{z}^{z}=\frac{m\left(m+n+p-1\right)}{t^{2}}=0,\quad R_{t}^{t}=\frac{m^{2}+n^{2}+p^{2}-m-n-p}{t^{2}}=0,
\end{equation}
leading to the following 4 solutions: 
\begin{eqnarray}
S_{\pm1}^{(GR)} & = & \left\{ m,n=\frac{1-m}{2}\pm\frac{\sqrt{Z}}{2},p=\frac{1-m}{2}\mp\frac{\sqrt{Z}}{2}\right\} ,\qquad Z=-3m^{2}+2m+1,\label{S1GR}\\
S_{2}^{(GR)} & = & \{m=0,n=0,p=1\},\qquad S_{3}^{(GR)}=\{m=0,n=0,p=0\}
\end{eqnarray}
with $S_{\pm1}^{(GR)}$ the best known solution and $S_{3}^{(GR)}$
denoting Milne spacetime $A=B=C=t$. Hence, the solution set contains
a sub-case vacuum model.

As mentioned before, condition $T_{abc}=0$ does not necessarily lead
to CG vacuum solutions for Bianchi I geometry. The most general energy
momentum tensor in the comoving frame $u^{a}=\delta_{b}^{a}$ for
\eqref{Bianchi1} is also a fluid with anisotropic pressure $T_{ab}=\rho(t)u_{a}u_{b}+p(t)h_{ab}+\Pi_{ab}$,
with the trace-less anisotropic pressure given as $\Pi_{b}^{a}=\hbox{diag}[0,P_{1},P_{2},-P_{1}-P_{2}]$
with $P_{1}(t),\,\,P_{2}(t)$ . Conservation of $T_{ab}$ implies:
\begin{equation}
\dot{\rho}=-(\rho+p)\left(\frac{\dot{A}}{A}+\frac{\dot{B}}{B}+\frac{\dot{C}}{C}\right)+\left(\frac{\dot{C}}{C}-\frac{\dot{A}}{A}\right)P_{1}+\left(\frac{\dot{C}}{C}-\frac{\dot{B}}{B}\right)P_{2},\label{BianchiF1}
\end{equation}
while condition $T_{abc}=0$ yields 
\begin{eqnarray}
3\dot{P}_{1}=(\rho+p)\left(\frac{\dot{C}}{C}+\frac{\dot{B}}{B}-\frac{2\dot{A}}{A}\right)-\left(\frac{\dot{C}}{C}+\frac{2\dot{A}}{A}\right)P_{1}-\left(\frac{\dot{C}}{C}-\frac{\dot{B}}{B}\right)P_{2},\label{BianchiF2}\\
3\dot{P}_{2}=(\rho+p)\left(\frac{\dot{C}}{C}-\frac{2\dot{B}}{B}+\frac{\dot{A}}{A}\right)-\left(\frac{\dot{C}}{C}-\frac{\dot{A}}{A}\right)P_{1}-\left(\frac{\dot{C}}{C}+\frac{2\dot{B}}{B}\right)P_{2},\label{BianchiF3}
\end{eqnarray}
Once an equation of state is selected (for example $p=p(\rho)$) the
system \eqref{BianchiF1}-\eqref{BianchiF3} can be integrated (3
variables $\rho,\,P_{1},\,P_{2}$ for three metric functions). It
is far from evident that $T_{ab}=0$ (from $C_{abc}=0$) holds.

To resolve the indeterminacy of vacuum solutions from $T_{ab}=0$,
we proceed as in the spherical vacuum case by using the Codazzi formulation
\eqref{Codazzi}-\eqref{preCodazzi} and following the Correspondence
Principle, in particular by relying on the same Kasner parametrization
that leads to a GR solution for vacuum Bianchi I. We obtain three
components 
\begin{eqnarray}
-\frac{\left(m-n\right)\left(2mn+mp+np-p^{2}-2m-2n-p+2\right)}{t^{3}}=0,\\
\frac{\left(n-p\right)\left(m^{2}-mn-mp-2np+m+2n+2p-2\right)}{t^{3}}=0,\\
\frac{\left(m-p\right)\left(mn+2mp-n^{2}+np-2m-n-2p+2\right)}{t^{3}}=0
\end{eqnarray}
However, since $g^{bc}C_{abc}=0$, only two of the three components
of $C_{abc}=0$ above are independent, hence we only consider two
components. We obtain the following solutions 
\begin{align}
S_{\pm1}^{(CG)} & =\left\{ m,n=\frac{1-m\pm\sqrt{Z}}{2},p=\frac{1-m\mp\sqrt{Z}}{2}\right\} ,\quad Z=-3m^{2}+2m+1,\label{S1CG}\\
S_{2}^{(CG)} & =\{m,n=m,p=1\},\,\,S_{3}^{(CG)}=\{m,n=m,p=m\}.\\
S_{4}^{(CG)} & =\left\{ m,n=\frac{2}{3},p=\frac{2}{3}\right\} .
\end{align}
Thus, the CG solution $S_{\pm1}^{(CG)}$ in \eqref{S1CG} reproduces
the known GR solution $S_{\pm1}^{(GR)}$ in \eqref{S1GR}. The CG
solutions $S_{2}^{(CG)},\,\,S_{3}^{(CG)}$ generalize GR solutions
$S_{2}^{(GR)},\,\,S_{3}^{(GR)}$, while $S_{4}^{(GR)}$ is not contained
in GR. Therefore, condition $g^{bc}C_{abc}=0$ did not restrict the
solution space of CG in Bianchi I vacuum. As expected, there are more
Bianchi I vacuum solutions in CG than in GR, but CG has a well defined
GR limit.

\section{FLRW models in CG}

Clement and Noiucer argue that FLRW models emerge in CG only as trivial
solutions. This is true only in the Cotton formulation \eqref{CottonFeqs}-\eqref{Cotton}
because it requires computing beforehand the Cotton tensor $C_{abc}$,
which vanishes automatically for all conformally flat spacetimes \cite{CottonWeyl}.
In fact, proceeding through \eqref{CottonFeqs}-\eqref{Cotton} leads
to an inconsistency ``$0=T_{abc}$'' for $T_{abc}\ne0$ in \eqref{CottonFeqs}.
Therefore, the Codazzi formulation \eqref{Codazzi}-\eqref{preCodazzi}
is the only way to examine FLRW models in CG. 
\begin{description}
\item [{Energy momentum tensors. }] In discussing the ``test cases''
of static spherical and Bianchi I vacuum with the Codazzi formulation
we presented the Codazzi formulation together with considering as
field source the same energy momentum tensor $T_{ab}$ as in GR instead
of $T_{abc}$ in \eqref{CottonFeqs}. However, CG under the Codazzi
formulation is also compatible with Einstein's equations in which
the right hand side of the Einstein tensor is an ``effective'' energy
momentum tensor. Assuming that ${\cal C}_{ab}$ is a Codazzi tensor
that solves \eqref{Codazzi}, we can express CG field equations as
\begin{equation}
G_{ab}=8\pi\,[T_{ab}+T_{ab}^{\hbox{\tiny{(eff)}}}],\qquad8\pi T_{ab}^{\hbox{\tiny{(eff)}}}={\cal C}_{ab}-\mathcal{C}\,g_{ab},\label{effT}
\end{equation}
We remark that usage of ``effective'' Einstein's equations as in
\eqref{effT} is very common (and useful) in other alternative theories,
for example in $f(R)$ theories, for which the field equations can
be written as \eqref{effT} \cite{Odin1,Odin2}: 
\begin{eqnarray}
G_{ab}=8\pi\left[\frac{T_{ab}+\frac{1}{16\pi}\left[(f(R)-R\,F(R)-2\Square F(R))g_{ab}+2\nabla_{a}\nabla_{b}F(R)\right)}{F(R)}\right],
\end{eqnarray}
where $T_{ab}$ is the energy momentum of GR, $R$ is the Ricci scalar,
$F(R)=\partial f(R)/\partial R$ and $\Square$ is the D'Alembertian
operator. Although this is much more complicated in $f(R)$ than in
CG, it is still useful because of the familiarity with Einstein's
equations, though it requires a proper interpretation of the geometric
terms. We summarize below CG solutions for FLRW models examined with
$T_{ab}$ and with $T_{ab}+T_{ab}^{\hbox{\tiny{(eff)}}}$.

\item [{FLRW models with $T_{ab}$.}] In \cite{Sussman1,Sussman2} Sussman
and Najera considered FLRW models characterized in GR by 
\begin{equation}
ds^{2}=-dt^{2}+a^{2}(t)\left[\frac{dr^{2}}{1-kr^{2}}+r^{2}(d\theta^{2}+\sin^{2}\theta d\phi^{2})\right],\label{flrw}
\end{equation}
with $T^{ab}=\rho u^{a}u^{b}+ph^{ab}\,\,\ ,u^{a}=\delta_{t}^{a}$
and $u_{a}\nabla_{b}T^{ab}=\dot{\rho}+\frac{3\dot{a}}{a}(\rho+p)=0$.
To apply \eqref{Codazzi}-\eqref{preCodazzi} to \eqref{flrw} without
modifying $T_{ab}$, the only possibility is to obtain ${\cal C}_{ab}\ne0$
by modifying the Friedman equation as 
\begin{equation}
H^{2}=\frac{\dot{a}^{2}}{a^{2}}=\frac{8\pi}{3}\rho-\frac{k}{a^{2}}\quad\hbox{extended\,to}\quad H^{2}=\frac{\dot{a}^{2}}{a^{2}}=\frac{8\pi}{3}\rho-\frac{k+\lambda{\cal K}(a)}{a^{2}},\label{flrw4}
\end{equation}
where ${\cal K}(a)$ is an arbitrary function and $\lambda$ is a
constant with 1/$\hbox{length}^{2}$ units. This new function is a
degree of freedom introduced by CG that modifies the time evolution
of the scale factor $a(t)$. It also modifies the extrinsic curvature
of the hypersurfaces orthogonal to the 4-velocity and their spatial
curvature given by their 3-dimensional Ricci scalar 
\begin{equation}
^{3}R=\frac{6(k+\lambda{\cal K})}{a^{2}},
\end{equation}
However, CG provides no evolution or conservation law for ${\cal K}$,
so to keep this CG solution compatible with a well posed initial value
problem we need to restrict it to be either constant or ${\cal K}=a^{2}$,
leading to a modified Friedman equation in \eqref{flrw4} that is
operationally the same as an FLRW model of GR with $\lambda=-(8\pi/3)\Lambda$.
However, there is an important conceptual difference: the kinematic
effect of a cosmological constant (accelerated expansion) comes from
spatial curvature, not from some $\Lambda$ approximating a dark energy
source. The modified Friedman equation in \eqref{flrw4} for a dust
source $\rho=\rho_{0}/a^{3}$ (representing cold dark matter) and
$k=0,\,\lambda<0$ takes the form: 
\begin{equation}
H^{2}=\frac{\dot{a}^{2}}{a^{2}}=\frac{8\pi}{3}\frac{\rho_{0}}{a^{3}}+|\lambda|,\label{flrw6}
\end{equation}
which is identical to the Friedman equation of the $\Lambda$CDM model
in GR if $|\lambda|=(8\pi/3)\Lambda$, except that the $\Lambda$CDM
model in CG acquires the covariant characterization as the unique
FLRW dust model with negative constant spatial curvature. 
\item [{FLRW models with an effective $T_{ab}$. }] Mantica and Molinari
\cite{Mantica2} have examined FLRW with an effective energy momentum
tensor of the form \eqref{effT} for various assumptions on ${\cal C}_{ab}$.
Applying \eqref{effT} for the FLRW metric allows the parametrization
of their Friedman equation to reproduce with a single free function
the Friedman equations of several alternative gravity theories, such
as $f(R)$, modified Gauss-Bonnet $f(G)$, teleparallel $f(T)$, including
the recently proposed Conformal Killing gravity \cite{Mantica2} and
Mimetic gravity. Mantica and Molinari relate these Friedman equations
to a Codazzi tensor that may be associated with dark sector sources.
It can also accommodate varying dark energy equations of state, such
as that by Chevallier-Polarski-Lindler model.\\
 The Codazzi condition for the effective (dark) stress-energy tensor
$8\pi T_{ab}^{\hbox{\tiny{({\rm eff)}}}}={\cal C}_{ab}-\mathcal{C}\,g_{ab}$
fixes or strongly constrains the geometry of space-time. Consider
the dark perfect-fluid stress-energy tensor $T_{ab}^{\hbox{\tiny{(eff)}}}=(\rho_{{\rm eff}}+p_{{\rm eff}})u_{a}u_{b}+p_{{\rm eff}}g_{ab}$.
In absence of acceleration, Theorem 2.1 in \cite{Mantica1} holds
(with obvious variation of notation):\\

\textbf{Theorem:} {\em The perfect fluid tensor ${\cal C}_{ab}=(\rho_{{\rm eff}}+p_{{\rm eff}})u_{a}u_{b}+\frac{1}{3}\rho_{{\rm eff}}g_{ab}$
is Codazzi if and only if: 
\begin{gather}
\nabla_{a}u_{b}=H(g_{ab}+u_{a}u_{b})\quad\text{with}\quad\;\nabla_{a}H=-\dot{H}u_{a}\label{eq: conditions GRW}\\
\nabla_{a}\rho_{{\rm eff}}=-\dot{\rho}_{{\rm eff}}u_{a},\quad\nabla_{a}p_{{\rm eff}}=-\dot{p}_{{\rm eff}}u_{a},\quad H=-\frac{\dot{\rho}_{{\rm eff}}}{3(\rho_{{\rm eff}}+p_{{\rm eff}})}=\frac{1}{3}\nabla_{a}u^{a},\label{HUBBLEGRW}
\end{gather}
The Codazzi perfect fluid tensor secures the space-time (because of
(\ref{eq: conditions GRW}) ) as generalised Robertson-Walker that
becomes (\ref{flrw}) in the conformally flat case.}.
\begin{rem}
The resulting Friedmann equations (eq (20) and (21)
in \cite{Mantica2}) have a further degree of freedom with respect
to GR. As stated in the introduction of \cite{Carloni 14} degeneracy
appears as additional unknown functions in extension of general relativity.
These are usually fixed using reconstructions methods \cite{Carloni 12}.
For example the Ricci scalar in $f(R)$ gravity is dynamically determined
by the trace of the stress-energy tensor via the equation $F(R)R-2f(R)+3\Square F(R)=8\pi T$
that possess non-trivial propagating solutions in vacuum: this suggests
the existence of a further degree of freedom (related to $R)$ with
respect to GR., as pointed out in \cite{Ferraro 18}. In this way
CG is not different.
\end{rem}

\subsection{Mimetic gravity }

Consider the field equations of Mimetic gravity (see
for example \cite{Sebastiani 17}): they are written in the form

\begin{equation}
R_{ab}-\frac{1}{2}Rg_{ab}=8\pi [T_{ab}+2\lambda u_{a}u_{l}+g_{ab}V(\phi)]\label{eq:Mimetic final}
\end{equation}
 where $V$ is a potential function and $\lambda$ is the Lagrange
multiplier and the relation $\nabla^{b}T_{ab}=0$ may be written as
$\nabla_{a}V=-2\dot{\lambda}u_{a}-6H\lambda u_{a}$: it implies that
$H=\frac{\dot{V}-2\dot{\lambda}}{6\lambda}$.We note that (\ref{eq:Mimetic final})
has the same form of (\ref{HUBBLEGRW}) with the identifications $p_{D}=V(t),\mu_{D}=\lambda(t)-2V(t)$.
Thus we claim 

\textcolor{red}{\medskip{}
}

\textbf{\textsc{In an FLRW space-time Cotton Gravity
has the same degree of freedom of mimetic gravity }}

\medskip{}
 Different choices of the potential $V$ give rise to several cosmological
models. On the other hand in \cite{Cardenas21} the specific form
of the potential was reconstructed using the Cevallier-Polarski-Linder
parametrization of $w$ as a function of the redshift $z$. The same
expression (eq 21 in \cite{Cardenas21}) is the eq 88 in \cite{Mantica2}.
\end{description}

\section{More general CG solutions do not increase ambiguity}

One of the claims of Clement and Noiucer is that CG becomes more problematic
as the symmetries of spacetimes became weaker. This might be a reasonable
guess with the Cotton formulation (though they only explored two test
case spacetimes). However, this claim is completely mistaken when
proceeding with the Codazzi formulation, as we have argued throughout
this article. In fact, this claim can be easily disproven by providing
as counterexamples of inhomogeneous CG solutions found in \cite{Mantica1}
and \cite{Sussman2}: the spherically symmetric Lemaitre-Tolman-Bondi
(LTB) dust solution, Szekeres models of class I, spherically static
perfect fluids and the Stephani Universe and its generalizations.

In none of these new CG solutions we found the confusion and ambiguities,
or ``non-predictive'' nature, reported by Clement and Noiucer. In
fact, the CG solutions we have found are self-consistent and exhibit
expected correspondence between them. As an example, we show below
that the Schwarzschild analogue of CG (\eqref{CGtrial2}) is the vacuum
limit of the CG analogue of LTB dust models, just as Schwarzschild
solution of GR is the vacuum limit of LTB models in GR \cite{kras}.

LTB solutions of GR for a dust source $T_{ab}=\rho u_{a}u_{b}$ and
$u^{a}=\delta_{t}^{a}$ are given by %
\begin{eqnarray}
ds^{2} & = & -dt^{2}+\frac{R'^{2}}{1-K}dr^{2}+R^{2}\,\left(d\theta^{2}+\sin^{2}\theta d\phi^{2}\right),\qquad2M'=8\pi\rho R^{2}R',\label{LTB1}\\
\dot{R}^{2} & = & \frac{2M(r)}{R}-K(r),\label{LTB2}
\end{eqnarray}
where $R=R(t,r)$, dot is $\partial/\partial t$ and primes denote
$\partial/\partial r$, $M(r)$ is the quasilocal Misner-Sharp mass.
Setting $M'=0$, so that $M=M_{s}=$ constant leads to $\rho=0$ in
\eqref{LTB1}, with the LTB solution becoming the Schwarzschild solution
in coordinates $(t,r)$ comoving with its timelike radial geodesics,
with proper time $t$ and each geodesic marked by $r$ (see page 299
of \cite{kras}). This comoving geodesic representation can be brought
to the standard static form \eqref{staticSS} in coordinates $(T,R)$
with $-A=B=1-2M_{s}/R$ by the inverse of this coordinate transformation
\begin{eqnarray*}
dT=-\sqrt{1-K}\,\left(1-\frac{2M_{s}}{R}\right)^{-1}dt+\left[\frac{\partial T}{\partial r}\right]_{t}dr,\qquad dR=\left(\frac{2M_{s}}{R}-K\right)^{1/2}dt+\left[\frac{\partial R}{\partial r}\right]_{t}dr,
\end{eqnarray*}
The LTB solution of CG derived in \cite{Sussman2} is described by
exactly the same metric and density equation as the LTB models of
GR in \eqref{LTB1}, but its equation \eqref{LTB2} for $\dot{R}$
is modified 
\begin{equation}
\dot{R}^{2}=\frac{2M(r)}{R}-K(r)-\gamma(r)\,R-\lambda(r)\,R^{2}.\label{LTBCG}
\end{equation}
The limit $M(r)=M_{s}$ and setting $\gamma(r)$, $\lambda(r)$ to
constants lead to the Schwarzschild (or Schwarzschild-Kottler) CG
analogues in \eqref{staticSS} and \eqref{Schw2}. This shows that
CG preserves the correspondence that holds for GR between a spherically
symmetric dust solution (LTB) and its spherical vacuum limit (Schwarzschild)
and is a strong argument of consistency.

\section{Conclusion}

We have provided extensive and solid argumentation to address the
critique by Clement and Noiucer of CG \cite{Clement}. Our response
in this article has emphasized the conceptual and practical limitations
of the Cotton formulation of the theory embodied in the field equations
\eqref{CottonFeqs}-\eqref{Cotton}, illustrating the advantages of
the Codazzi formulation that is better suited to address the concerns
raised by these authors. Their exclusive usage of the Cotton formulation
explains the confusion and errors found in  \cite{Clement}. We also emphasize that we are aware of the
limitations of CG, even with the approach we have described. CG must
be tested, criticized and examined further, but not dismissed. 
It is also important to consider that none of the proposed alternative
gravity theories has managed to achieve a satisfactory extension or
modification of GR.

\end{document}